\documentclass[aps,prl,superscriptaddress,floatfix,amsmath,amsfonts,twocolumn]{revtex4-2}  
\usepackage[T1]{fontenc}
\usepackage[utf8]{inputenc}
\usepackage{amsmath,amssymb,amsfonts,bm}
\usepackage{graphicx}
\usepackage{hyperref}
\usepackage[dvipsnames]{xcolor}

\begin{document}

\title{Synchronization-induced flat bands in driven-dissipative dimer-waveguide chains}

\author{A.~N.~Osipov}
\email{alexey.n.osipov@gmail.com}
\affiliation{Technion – Israel Institute of Technology, 32000 Haifa, Israel}
\affiliation{Department of Physics, Guangdong Technion–Israel Institute of Technology, 241 Daxue Road, Shantou, Guangdong 515063, China}
\affiliation{Department of Physics, ITMO University, Saint~Petersburg 197101, Russia}

\author{I.~G.~Savenko}
\affiliation{Department of Physics, Guangdong Technion–Israel Institute of Technology, 241 Daxue Road, Shantou, Guangdong 515063, China}
\affiliation{Technion – Israel Institute of Technology, 32000 Haifa, Israel}

\author{Sergej Flach}
\affiliation{Institute for Basic Science, Daejeon 34126, Republic of Korea}
\affiliation{Basic Science Program, Korea University of Science and Technology, Daejeon 34113, Republic of Korea}
\affiliation{Centre for Theoretical Chemistry and Physics, The New Zealand Institute for Advanced
Study (NZIAS), Massey University Albany, Auckland 0745,
New Zealand}

\author{A.~V.~Yulin}
\affiliation{Department of Physics, ITMO University, Saint~Petersburg 197101, Russia}

\date{\today}

\begin{abstract}
Flat bands in driven-dissipative systems offer a route to engineer strongly localized, long-lived excitations, yet their selective population via incoherent pumping remains an open challenge. 
We study a one-dimensional chain of coupled lasing dimers arranged in a cross-stitch geometry and show that the synchronization regime of the individual dimers, controllable through pump intensity or inter-resonator distance, determines the character of the flat band hosted by the chain. In the in-phase (ferromagnetic) regime, the flat band appears as a subdominant, damped mode in the linear excitation spectrum. In the antiphase (antiferromagnetic) regime, by contrast, the dimers decouple and the flat band becomes the dominant, neutrally stable mode: it corresponds to an infinite family of Goldstone modes arising from the independent phase rotations of non-interacting dimers, and its compact localized states are directly observable in the noise response spectrum. Switching between these two regimes via pump control constitutes a pump-induced phase transition of the lasing lattice. Our results establish synchronization engineering as a practical mechanism for selective flat-band population in driven-dissipative optical systems, and open new avenues for studying flat-band physics, including nonlinear effects, Fano resonances, and excitation coherence in experimentally accessible laser and polariton platforms.
\end{abstract}


\maketitle

\textit{\textcolor{blue}{Introduction.---}}
Flat bands have been a central focus of the scientific community for approximately forty years~\cite{leykam2018artificial,Leykam:2018aa,danieli2024flat,danieli2026progressartificialflatband}, dating back to the first theoretical treatments of the subject~\cite{PhysRevB.34.5208, PhysRevLett.62.1201}. The emergence of a flat band in a physical system leads to a variety of unusual properties, including the localization of excitations~\cite{PhysRevLett.81.5888, PhysRevB.88.224203}, singularities in the density of states, enhanced interaction effects~\cite{tsai2015interaction}, and the appearance of exotic quantum~\cite{PhysRevB.85.085209, PhysRevA.88.063613} and topological states~\cite{PhysRevLett.129.253001, PhysRevB.96.205304}. 

However, despite their potential to host a wealth of intriguing phenomena, flat bands are highly sensitive to perturbations and typically rely on either symmetry protection or precise fine-tuning of system parameters. This sensitivity poses significant challenges for their experimental observation and manipulation. Nevertheless, recent advances in experimental techniques have begun to overcome these obstacles, enabling the excitation and probing of flat bands across a variety of platforms. Flat bands have been experimentally demonstrated in electronic lattices~\cite{lape2025realization}, cold atoms~\cite{taie2015coherent, aidelsburger2015measuring}, photonic crystals~\cite{vicencio2015observation}, and polaritonic condensates~\cite{baboux2016bosonic, klembt2017polariton, alyatkin2021quantum}. The latter system offers a high degree of tunability for optical excitation~\cite{whittaker2018exciton, Sun2018Excitation, ko2020partial}, allowing for the dynamical compensation of sample imperfections by adjusting the pump intensity~\cite{alyatkin2021quantum, johnston2024macroscopic}.

It is also instructive to draw a connection between flat-band systems and bound states in the continuum (BICs), another topic of significant recent interest. The BIC phenomenon arises when radiation from the system of resonators is suppressed due to the destructive interference of fields emitted by individual resonators~\cite{hsu2016bound, plotnik2011experimental, bulgakov2008bound}. Both flat bands and BICs are closely related to practical concepts in photonics, such as slow-light generation~\cite{Krauss2007}, high-quality lasers~\cite{Hirose_2014}, and chemical or biological sensing~\cite{Yanik2011}. Despite this progress, selective population of flat bands in realistic driven-dissipative optical systems under incoherent pumping remains an open challenge.

In this Letter, we develop an approach applicable to a wide range of physical systems that provides a generic solution for the controlled manipulation of waves within flat bands. This offers a new perspective on populating flat bands and understanding flat band excitations. We consider flat bands in a general class of lasing systems composed of a lattice of coupled active optical resonators. Meanwhile, we extend the study of flat bands to non-Hermitian setups, where optical waves propagating along a chain can exhibit dispersions featuring flat bands. In these systems, the pump compensates for losses, enabling the existence of non-trivial stationary states.

\begin{figure}[!t]
\centering
\includegraphics[width=\linewidth]{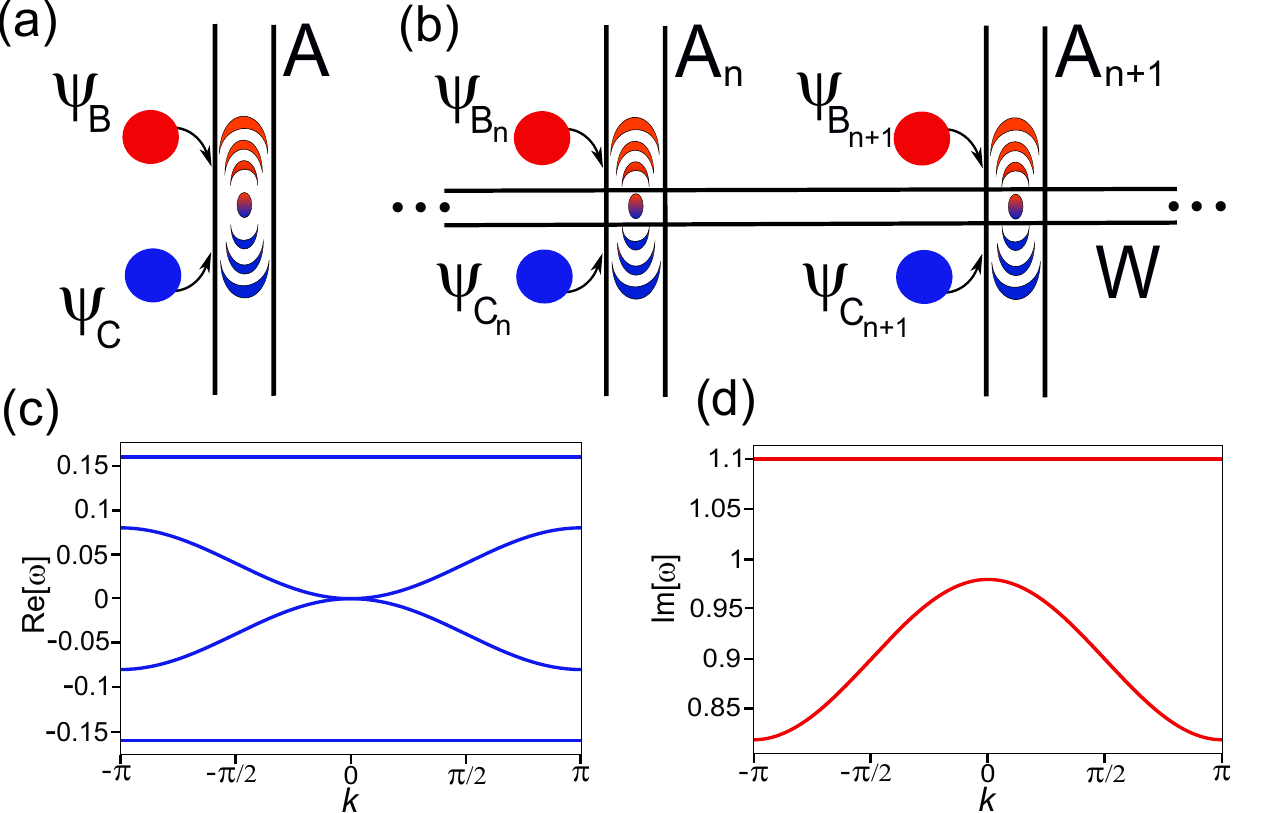}
\caption{(a) Schematic illustration of a single laser dimer consisting of two lasing resonators described by order parameters $\psi_{B}$ and $\psi_{C}$, coupled to a waveguide $A$. A standing wave inside the waveguide has a node at the center, corresponding to the antisymmetric configuration $\psi_C = -\psi_B$. (b) Schematic illustration of a chain of laser dimers coupled to a common horizontal waveguide $W$. Real (c) and imaginary (d) parts of the linear excitation spectrum derived from Eq.~\eqref{chain equation} with respect to the trivial solution, computed for the following parameters: $J=-0.1$, $P=2$, $g=1$, $\alpha = 1$, $\kappa = -1$, $|\zeta| = 0.02$, $\chi = -0.5$.}
\label{fig_1}
\end{figure}

To be specific, we model an array of exciton-polariton lasers coupled via optical waveguides as a non-Hermitian flat-band system (Fig.~\ref{fig_1}a). These units, which we refer to as ``laser dimers,'' represent the fundamental building blocks of the system. They interact through optical excitations propagating in a horizontal waveguide, as shown in Fig.~\ref{fig_1}(b). Consequently, the system supports wave propagation along the array. The symmetry of the configuration ensures that antisymmetric modes do not excite waves in the horizontal waveguide, rendering these states compact. The existence of such compact localized states (CLSs) implies that the dispersion relation contains flat bands. Importantly, as we demonstrate below, this property also holds for linear excitations on top of a spatially uniform nonlinear background.

We focus on the regime where the interaction between lasers can be understood in terms of \textit{synchronization}---a fundamental concept that illuminates many aspects of the coherence-preserving evolution in driven-dissipative systems. This approach has been effectively employed to analyze the behavior of polariton condensate arrays emulating the XY model~\cite{berloff2017realizing}, exhibiting geometric frustration~\cite{nixon2013observing, cookson2021geometric}, and demonstrating spontaneous symmetry breaking~\cite{johnston2021artificial, dolinina2022spontaneous}.

\textit{\textcolor{blue}{Model.---}}
The dynamics of the system is governed by a set of partial differential equations (PDEs) for the fields $A_i$ and $W$ in the waveguides, coupled with ordinary differential equations (ODEs) for the laser mode amplitudes $\psi_{B_i,C_i}$. We assume that the field distribution in the lasers (i.e., the laser modes) remains largely unperturbed by the coupling to the waveguides, as well as by dissipative and nonlinear effects. Consequently, the lasers are characterized by the complex amplitudes of their respective lasing modes:
\begin{subequations}
\label{condensate equation}
\begin{align}
&\partial_t \psi_{B_i, C_i}(t) = H_0(\psi_{B_i, C_i}) + \sigma A_i(x_{B_i, C_i}),\\
&\partial_t A_i(t,x) = -\gamma A_i + \hat{D}A_i + \sigma\sum_{B_i, C_i} \delta(x_{B_i, C_i})\psi_{B_i, C_i}\nonumber \\
&~~~~~~~~~~~~~~~~+\eta\delta(x_0)W(t, y_i), \\
&\partial_t W(t,y) = -(\Gamma+i\omega_W) W + \hat{D}_W W + \eta\sum_{i} \delta(y_i)A_i(x_0).
\end{align}
\label{Coupled dimers WG dynamic equations}
\end{subequations}
We now define each term in Eq.~\eqref{Coupled dimers WG dynamic equations} individually. 

First, in the absence of coupling, the evolution of an isolated laser is described by $\partial_t\psi = H_0(\psi)$, where:
\begin{align}
\label{H0}
H_0(\psi) = -i\omega_0 \psi + \left((1-ig)\frac{P}{1+|\psi|^2}-1\right)\psi - i\alpha|\psi|^2\psi.
\end{align}
Here, $P$ is the normalized pumping intensity, $g$ is the reservoir-induced blueshift, $\alpha$ is the particle-particle interaction strength, and $\omega_0$ is the detuning of the free condensate frequency relative to the cutoff frequency of the waveguide mode. Equation~\eqref{H0} serves as a tight-binding approximation of the Gross-Pitaevskii equation for exciton-polariton condensates~\cite{PhysRevLett.99.140402} or, alternatively, describes an atomic laser~\cite{PhysRevA.58.4841} with an adiabatically eliminated reservoir.

Second, the free-field evolution in the waveguides is governed by the terms $-\gamma A_i + \hat{D}A_i$ and $-(\Gamma+i\omega_W) W + \hat{D}_W W$. The parameters $\gamma$ and $\Gamma$ represent the damping rates for the vertical and horizontal waveguides, respectively, while $\omega_W$ denotes the detuning of the horizontal waveguide's cutoff frequency relative to the vertical one. The operators $\hat{D}$ and $\hat{D}_W$ describe wave dispersion. Near the cutoff frequency, the dispersion is approximated quadratically as $\hat{D} = \frac{i}{2} \partial_x^2$ and $\hat{D}_W = \frac{i}{2} \partial_y^2$, where the dimensionless coordinates $x$ and $y$ are normalized such that the coefficients are equal.

Third, we assume each laser couples exclusively to the nearest point on its vertical waveguide; thus, $A_i(x_{B_i, C_i})$ represents the field at the specific interaction point. In Eq.~\eqref{Coupled dimers WG dynamic equations}c, the horizontal waveguide field $W(t, y)$ interacts with the vertical waveguides at their midpoints $x_0$, located at positions $y_i$.

When the pump intensity exceeds the threshold value~\cite{Aleiner2012Radiative}, lasing sets in and the field amplitude is dictated by the energy balance. For isolated resonators, this condition implies $P/(1+|\psi|^2) = 1$, with a lasing frequency $\omega = \text{Im}[H_0(\psi)/\psi]$. We treat the coupling between resonators and waveguides perturbatively and assume that the delay time is short relative to the synchronization timescales. 
In this limit, the role of the waveguides is reduced to providing quasi-instantaneous couplings. 
As shown in the Supplementary Materials~\cite{[{See Supplementary Materials at [URL], which gives the details of the derivations}]SMBG}, the system reduces to a set of coupled ODEs:
\begin{subequations}
\label{chain equation}
\begin{align}
\partial_t \psi_{B_i} &= \left(-i\omega_0 +i\alpha |\psi_{B_i}|^2 + \frac{(1-ig)P}{1+|\psi_{B_i}|^2}-1\right)\psi_{B_i}  \\ \nonumber
&+ J(1 - i\varkappa) \psi_{C_i} + \sum_{\langle j \rangle}\zeta (1 - i\chi)(\psi_{B_j} + \psi_{C_j} ), \\
\partial_t \psi_{C_i} &= \left(-i\omega_0 + i\alpha |\psi_{C_i}|^2 + \frac{(1-ig)P}{1+|\psi_{C_i}|^2}-1 \right)\psi_{C_i}  \\ \nonumber
&+ J(1 - i\varkappa) \psi_{B_i} + \sum_{\langle j \rangle}\zeta (1 - i\chi)(\psi_{B_j} + \psi_{C_j} ).
\end{align}
\end{subequations}
The parameters $J = \operatorname{Re}[\sigma^2 e^{i\kappa L}/\kappa]$ and $\varkappa = -\operatorname{Im}[e^{i\kappa L}/\kappa]/\operatorname{Re}[e^{i\kappa L}/\kappa]$ characterize the effective intra-dimer coupling, where $\kappa = \sqrt{2\omega + 2i\gamma}$ is the wave vector at frequency $\omega$ and $L = |x_C - x_B|$. The inter-dimer coupling parameters $\zeta \sim e^{i\kappa_W L_W}$ and $\chi$ depend on the wave vector $\kappa_W = \sqrt{2(\omega-\omega_W) + 2i\Gamma}$ and the spacing $L_W = |y_n-y_{n-1}|$. Given that the coupling scales exponentially with distance, we consider only nearest-neighbor interactions ($\langle j \rangle$) and assume $\zeta \ll J$~\cite{SMBG}.

The symmetry of these inter-dimer interactions resembles a cross-stitch lattice. Our calculations of the linear excitation spectra for $|\psi| \ll 1$ confirm that the dispersion contains both dispersive and flat bands, where the eigenfrequencies are independent of the wavenumber $k$ (Figs.~\ref{fig_1}c, d). At the lasing threshold ($P \approx 0.9$ for our parameters), the imaginary part of one eigenfrequency becomes positive. For $J < 0$, the fastest-growing mode resides in the flat band, which subsequently suppresses other modes and defines the symmetry of the nonlinear lasing state.

\textit{\textcolor{blue}{Lasing dynamics of an isolated dimer.---}}
By neglecting the inter-dimer coupling in Eq.~\eqref{chain equation}, we arrive at the dynamical equation for an isolated laser dimer:
\begin{align}
\label{EqCoupledModel01}
\partial_t\psi_{B,C} = &-i\omega_0 \psi_{B,C}+ \left((1-ig)\frac{P}{1+|\psi_{B,C}|^2}-1\right)\psi_{B,C} \nonumber \\
&- i\alpha|\psi_{B,C}|^2\psi_{B,C} +J(1-i\varkappa)\psi_{C, B}.
\end{align}
In the weak coupling regime, we can adopt the Kuramoto approximation and characterize the laser modes $\psi = \sqrt{\rho} e^{i\phi}$ primarily by their phases $\phi$. This reduction is valid for relatively high pump intensities, where the lasing densities $\rho = |\psi|^2$ are strongly clamped by the energy balance. The equation governing the phase difference $\Delta \phi = \phi_B - \phi_C$ is given by~\cite{SMBG}:
\begin{align}
\label{Kuramoto approximation}
\partial_t (\Delta \phi) = -J(1-\alpha \varkappa P + g \varkappa) \sin(\Delta \phi).
\end{align}
Evidently, the model possesses equilibrium points at $\Delta\phi=0$ and $\Delta\phi=\pi$. The dynamical stability of these states is determined by the sign of the effective coupling $J_{\text{eff}} = J(1-\alpha \varkappa P + g \varkappa)$. Consequently, in the stationary regime, the lasers within a dimer synchronize either in-phase or in anti-phase.

To quantify the synchronization, it is instructive to introduce the parameter $S_1 = 2\operatorname{Re}[\psi_B\psi_C^*]$, which serves as a direct analog of the first Stokes parameter describing the linear polarization of an electromagnetic field. An in-phase solution ($\Delta\phi=0$) corresponds to $S_1 = 1$, and our analysis predicts its stability when $J_{\text{eff}}>0$. Conversely, the anti-phase solution ($\Delta\phi=\pi$) yields $S_1 = -1$ and is stable when $J_{\text{eff}}<0$. It should be noted that both $J$ and $\varkappa$ depend on the laser frequency and the inter-laser distance.

When $J_{\text{eff}} \approx 0$, the first-order approximation in Eq.~\eqref{Kuramoto approximation} becomes insufficient, necessitating a higher-order expansion to accurately describe the dynamics~\cite{SMBG}. In such cases, asymmetric equilibrium points may emerge via a pitchfork bifurcation~\cite{johnston2021artificial, dolinina2022spontaneous,SMBG}.

\begin{figure}[!tb]
\centering
\includegraphics[width=0.9\linewidth]{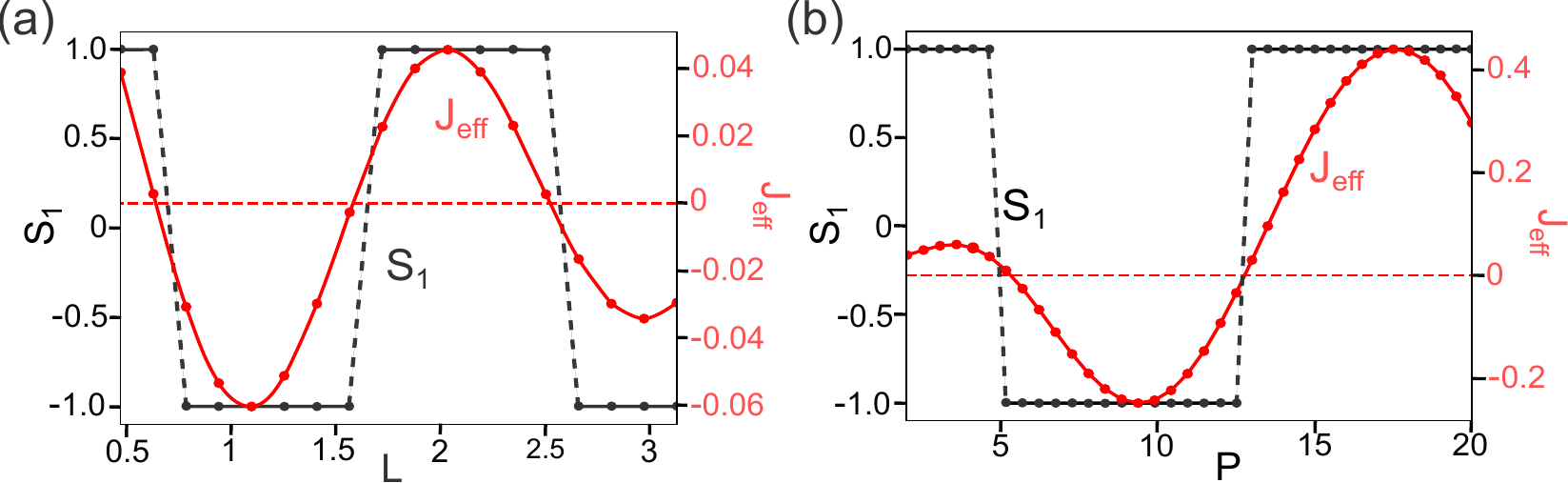}
\caption{Synchronization regimes in a single laser dimer. 
(a) The Stokes parameter $S_1 = 2\operatorname{Re}[\psi_B\psi_C^*]$ from Eq.~\eqref{condensate equation} as a function of the inter-spot distance $L$ at fixed pump intensity $P=2$ (black solid curve). The values $S_1 = \pm1$ correspond to in-phase ($\Delta\phi=0$) and anti-phase ($\Delta\phi=\pi$) synchronization, respectively. Red solid curve represents the effective coupling parameter $J_\textrm{eff}$ derived from Eq.~\eqref{Kuramoto approximation}. 
(b) $S_1$ and $J_\textrm{eff}$ as functions of the pump intensity $P$ for a fixed inter-spot distance $L\approx 2.2$. Other parameters of the model are listed in~\cite{Sim_parameters}.
}
\label{fig_2}
\end{figure}

Figures~\ref{fig_2}(a) and (b) show the function $J_{\text{eff}}$ plotted alongside the numerically obtained stationary parameter $S_1$ (from Eq.~\eqref{condensate equation}) as a function of the distance between resonators and as a function of pump intensity, respectively. These results confirm that the sign of $J_{\text{eff}}$ dictates the relative phase of the lasers. While the stability switching points are slightly shifted from the theoretical predictions, the analytical model exhibits excellent agreement with numerical simulations. A key conclusion from Fig.~\ref{fig_2} is that the symmetry of the stationary state can be controlled by either varying the inter-condensate distance or adjusting the pump intensity, the latter being particularly convenient for experimental implementation.

\textit{\textcolor{blue}{Collective dynamics and linear excitation spectra in a dimer chain.---}}
Let us now turn to the interaction between dimers. To understand the stability and dynamics of the chain in the lasing regime, we analyze the spectrum of linear excitations on top of the stationary background. The character of these excitations depends drastically on the states of individual dimers.

For the trivial solution (vanishing field), the linear excitation spectra exhibit both flat and dispersive bands regardless of the sign of $J_{\text{eff}}$. However, the gain distribution differs markedly between the two cases: when $J_{\text{eff}} < 0$, the flat bands possess higher gain, whereas for $J_{\text{eff}} > 0$, the situation is reversed (see Fig.~S3 in \cite{SMBG}). This asymmetry predicts that the nonlinear stage of evolution will be dominated by different modes depending on the synchronization regime.

Numerical simulations confirm that for parameters favoring the antisymmetric stable state of individual dimers ($\psi_{B_i} = -\psi_{C_i}$, corresponding to $J_{\text{eff}} < 0$), the stable solutions of the chain consist of a sequence of noninteracting non-synchronized dimers. Starting from random noise initial conditions, the system evolves into a state where each dimer acquires an independent random phase. 

When the dimers' phases are random, the equations governing elementary excitations are not translationally invariant, which complicates the analysis. Indeed, to study the band structure, translational invariance is required. To avoid this difficulty, we first consider excitations on top of a background with homogeneous dimer phases. Such a configuration can be achieved, for instance, by adiabatically switching the coupling sign $J_{\text{eff}}$ through pump tuning, starting from a ferromagnetically synchronized chain. Importantly, due to the system symmetry, the existence of flat bands in the excitation spectrum does not depend on the phase distribution and can be observed even when the nonlinear background lacks translational invariance (see \cite{SMBG} for more details). 

\begin{figure}[!tb]
\centering
\includegraphics[width=0.88\linewidth]{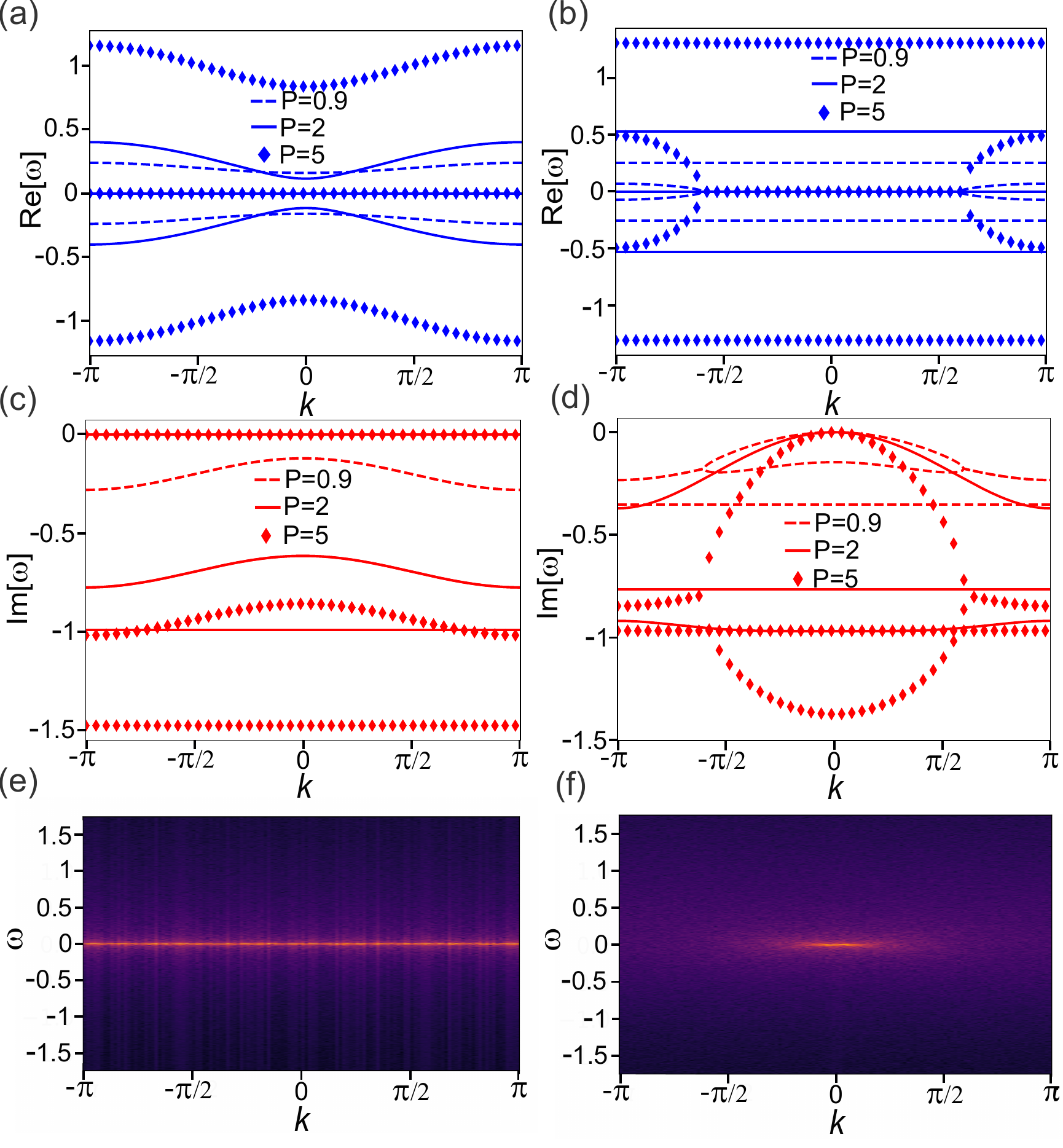}
\caption{Linear excitation spectra of the chain described by Eq.~\eqref{chain equation}. Real (a,b) and imaginary (c,d) parts of the modes as functions of the wave number $k$ for different pump intensities: weak lasing $P = 0.9$ (dashed lines), moderate pump $P = 2$ (solid lines), and stronger pump $P = 5$ (clubs). Panels (a) and (c) show the spectra for $J = -0.1$, corresponding to the dimers in the antisymmetric configuration ($\Delta\phi=\pi$). Panels (b) and (d) are for $J = 0.1$, corresponding to the dimers in the symmetric state ($\Delta\phi=0$). (e) Noise response spectrum for the chain of noninteracting antisymmetric dimers with equal phases. (f) Noise response spectrum for the chain of interacting phase-locked ferromagnetic dimers. We used $g=1$, $\alpha = 1$, $\kappa = -1$, $\zeta = 0.02$, $\chi = -0.5$.}
\label{fig3}
\end{figure}

Figures~\ref{fig3}(b) and \ref{fig3}(d) show the computed excitation spectra. Remarkably, the nonzero background does not destroy the flat bands but rather shifts their positions. Thus, the background modifies the dispersion characteristics of linear excitations. For the dispersive (non-flat) branches, the imaginary parts of the eigenfrequencies are negative, indicating that these excitations decay in time. Thus, the background is stable against perturbations belonging to these bands. In contrast, for the flat band, both the real and imaginary parts of the eigenfrequency are exactly zero. This implies that the background is neutrally stable against perturbations belonging to this flat band. Consequently, the excitation of such modes results in everlasting oscillations residing on the background. 
This observation can be interpreted as follows: due to the absence of coupling between dimers, the relative phase between them is not fixed. Each dimer can rotate its phase independently, giving rise to an infinite number of Goldstone modes. 

The spectrum of linear excitations reflects how the system responds to external perturbations. In particular, it determines the stability of a given solution, the noise response characteristics, the signal correlation and signal transmission properties. For instance, Fano resonances may arise due to the presence of flat bands~\cite{Ramachandran2018}. If a weak signal $f_i^{B,C}(t)$ is added to the right-hand side of Eq.~\eqref{chain equation}, the linear response can be obtained by solving a Langevin-type equation $\partial_t\xi = \hat{\mathcal{L}}\xi + \tilde{f}(t)$, where $\xi$ is the vector of elementary corrections, $\hat{\mathcal{L}}$ is the linearized operator~\cite{SMBG}, and $\tilde{f}(t)$ represents the external force. Knowledge of the spectrum of $\hat{\mathcal{L}}$ is therefore essential for describing the system's response to external signals. 

Here, we limit ourselves to considering only the response to weak white noise added as a stochastic driving force $f_i^{M}(t)$ to the right-hand side of Eq.~\eqref{chain equation}, satisfying $\langle f_i^{M}(t)f_j^{M'}(t')\rangle \sim \delta_{MM'}\delta_{ij}\delta(t-t')$, where $M = B, C$ labels the resonators within a dimer, and $i, j$ index the dimers along the chain. Figure~\ref{fig3}(e) shows the excitation spectrum of the chain composed of noninteracting antiphase dimers exposed to weak white noise. A flat band is clearly visible in the noise response spectrum. This band comprises the Goldstone modes with both zero real and imaginary parts of the eigenfrequency and represents the dominantly excited mode.

Let us now consider the case where individual dimers are synchronized in-phase (ferromagnetic state). Figure~\ref{fig3}(b,d) shows the linear excitation spectra for a chain of ferromagnetic dimers, computed on the background of the stable nonlinear phase-locked solution for the chain. In this configuration, flat bands also appear in the excitation spectrum. However, unlike the antiferromagnetic case, these flat-band modes now possess eigenfrequencies with negative real parts and therefore decay in time. 

A particularly interesting feature emerges in the dispersive bands when the nonlinear background is taken into account. While for the linear Hamiltonian these bands are purely dispersive in their real parts (see Fig.~\ref{fig3}(b)), the presence of a nontrivial background modifies their structure. For low pump intensities (dashed lines in Fig.~\ref{fig3}(b)), parts of these bands begin to flatten. At moderate pumping (solid lines), the real parts of the frequencies become completely flat, with one state becoming double-degenerate. As the pump increases further (clubs), the bands split again and regain their dispersive character. The only Goldstone mode in this case corresponds to a global phase rotation of the entire chain. This mode has $\operatorname{Im}[\omega] = 0$, indicating neutral stability, and is the most prominently excited mode in the noise response spectrum (Fig.~\ref{fig3}(f)).

It is worth noting that for sufficiently long chains, the stable configurations are not limited to purely phase-locked configurations of the dimer chain. Frustrated (vortex) solutions with a linearly growing phase along the chain can also emerge as stable states (see examples of such configurations in~\cite{SMBG}).

\begin{figure}[!tb]
\centering
\includegraphics[width=0.9\linewidth]{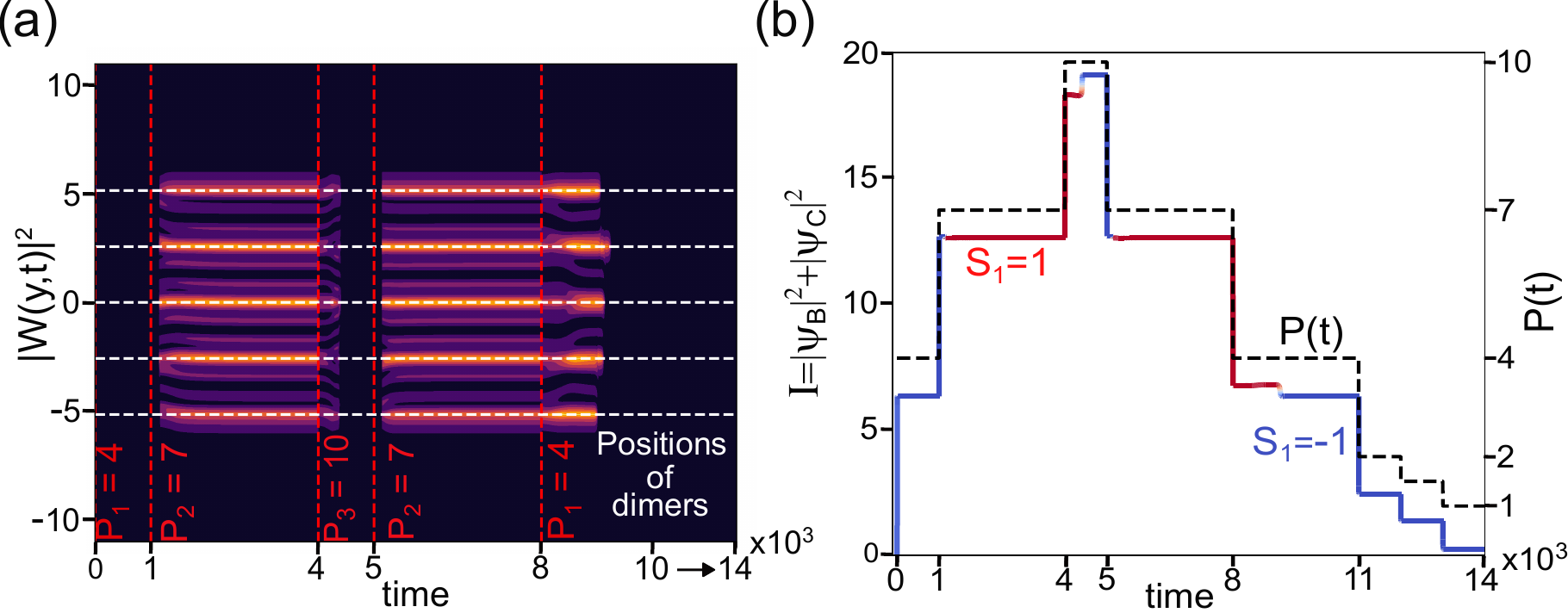}
\caption{Dynamics of the five-dimer chain under time-varying pump intensity. (a) Waveguide field intensity $|W|^2$ as a function of distance $y$ and time $t$. White dashed lines indicate the positions of the lasing dimers along the waveguide. (b) Total density $I(t) = |\psi_B(t)|^2 + |\psi_C(t)|^2$ of the central dimer as a function of time (the other dimers exhibit similar behavior). The color of the curve reflects the value of the Stokes parameter $S_1$: blue corresponds to $S_1 = -1$ (antiphase dimer configuration), while red corresponds to $S_1 = 1$ (in-phase dimer configuration). Black dashed line shows the time-dependent pump intensity, which changes abruptly at $t = [0, 1, 4, 5, 8, 11, 12, 13] \times 10^3$. The pumps $P = 4$, $P = 10$, and $P < 4$ correspond to $J_{\text{eff}} < 0$, whereas $P = 7$ corresponds to $J_{\text{eff}} > 0$. The parameters of the model are listed in~\cite{Sim_parameters}.
}
\label{fig4}
\end{figure}

Next, we perform numerical simulations of the system of coupled dimers described by Eqs.~\eqref{Coupled dimers WG dynamic equations} for a chain consisting of five coupled dimers. The inter-spot distance $L$ and frequency detuning $\omega_0$ are chosen to achieve antiferromagnetic coupling in dimers at low pump intensities. Starting from weak noise initial conditions, we apply a pump with intensity $P_1 = 4$. The system evolves for $t=1000$, which is a sufficient time for a stationary state to establish. The numerical simulations confirm that in this regime, no waves are excited in the horizontal waveguide due to the antisymmetric configuration of the dimers, which prevents radiation to the waveguide $W$ (Fig.~\ref{fig4}(a)).

Then, the pump is increased to $P_2 = 7$ and the simulation continues. At this higher pump, individual dimers transition to the ferromagnetic state, and consequently, they start to emit waves into the horizontal waveguide. The appearance of waveguide radiation is clearly visible in Fig.~\ref{fig4}(a). The synchronization regimes of individual dimers can be monitored through the Stokes parameter $S_1$, shown in Fig.~\ref{fig4}(b). The observed behavior is in good agreement with the Kuramoto approximation described in the previous section.

Subsequently, the increase of pumping to $P_3 = 10$ returns the dimers to the antisymmetric configuration and leads to the disappearance of waveguide radiation. After a stable state is reestablished, we lower the pump back to $P_2 = 7$. In the final stage of the simulation, we gradually decrease the pump intensity to $P = 1$ at $t = 13000$, thereby establishing a weak lasing regime in the system. At this pump level, the dimers remain in the antiferromagnetic state, and no radiation is observed in the waveguide.

As the pump decreases, the intensities of the dimer modes decay, and the system gradually enters the linear regime. Importantly, switching off the pump preserves the phase difference of the lasers in each dimer. Since the dimers were in the antiferromagnetic state prior to pump switch-off, the resulting linear state belongs to the flat band. This procedure thus provides a practical method for selectively exciting flat-band states. It is worth noting that for certain parameter ranges, bistability or oscillatory dynamics may occur; however, these phenomena lie beyond the scope of the present Letter.

\textit{\textcolor{blue}{In summary,}}
we have demonstrated that a chain of optically coupled lasing dimers provides a tunable and versatile platform for investigating flat-band physics within driven-dissipative systems. Our analysis shows that flat bands emerge in the linear excitation spectra regardless of the synchronization regime; however, their physical characteristics differ significantly. In the antiferromagnetic (antiphase) case, the flat bands manifest as Goldstone modes that dominate the noise response, whereas in the ferromagnetic (in-phase) case, these modes are dissipative and decay over time.

By controlling the non-resonant pump intensity, we can switch between these synchronization regimes, effectively inducing "phase transitions" in the lasing lattice. This tunability allows us to propose a robust scheme for the selective population of flat-band states. The procedure involves three primary stages: first, the dimer parameters are tuned to achieve asymmetric synchronization in the weak-lasing regime.
Then, the system is pumped above the lasing threshold, ensuring the antisymmetric configuration is the sole stable attractor.
Finally, once the stationary state is established, the pump is switched off, allowing the system to evolve into a quasilinear regime.
Because the symmetry of the nonlinear stable state (in the absence of inter-dimer coupling) matches that of the linear eigenmode of the flat band, the system persists as a populated flat-band state during the subsequent decay.

The ability to selectively populate and manipulate these flat-band excitations in driven-dissipative systems opens new avenues for exploring non-Hermitian phenomena. These findings have potential implications for the design of active photonic lattices, with applications in high-coherence signal transmission, controlled noise response, and the engineering of excitation coherence in large-scale laser arrays.

\textit{\textcolor{blue}{Acknowledgments---}}We were supported by the National Natural Science Foundation of China (NSFC) under Grant No.~W2532001.
SF acknowledges financial support from the Institute for Basic Science (IBS) in the Republic of Korea through Project No.
IBS-R024-D1 and IBS-R041-D1-2026-a00. The work of AY was supported by Russian Science Foundation, grant 23-72-00031.

\begin{acknowledgments}

\end{acknowledgments}

\bibliographystyle{apsrev4-2}
\bibliography{references}

@article{leykam2018artificial,
  title={Artificial flat band systems: from lattice models to experiments},
  author={Leykam, Daniel and Andreanov, Alexei and Flach, Sergej},
  journal={Advances in Physics: X},
  volume={3},
  number={1},
  pages={1473052},
  year={2018},
  publisher={Taylor \& Francis}
}

@article{PhysRevB.34.5208,
  title = {Localization of electronic wave functions due to local topology},
  author = {Sutherland, Bill},
  journal = {Phys. Rev. B},
  volume = {34},
  issue = {8},
  pages = {5208--5211},
  numpages = {0},
  year = {1986},
  month = {Oct},
  publisher = {American Physical Society},
  doi = {10.1103/PhysRevB.34.5208},
  url = {https://link.aps.org/doi/10.1103/PhysRevB.34.5208}
}

@misc{SMBG
}

@article{PhysRevLett.62.1201,
  title = {Two theorems on the Hubbard model},
  author = {Lieb, Elliott H.},
  journal = {Phys. Rev. Lett.},
  volume = {62},
  issue = {10},
  pages = {1201--1204},
  numpages = {0},
  year = {1989},
  month = {Mar},
  publisher = {American Physical Society},
  doi = {10.1103/PhysRevLett.62.1201},
  url = {https://link.aps.org/doi/10.1103/PhysRevLett.62.1201}
}

@article{PhysRevLett.81.5888,
  title = {Aharonov-Bohm Cages in Two-Dimensional Structures},
  author = {Vidal, Julien and Mosseri, R\'emy and Dou\ifmmode \mbox{\c{c}}\else \c{c}\fi{}ot, Benoit},
  journal = {Phys. Rev. Lett.},
  volume = {81},
  issue = {26},
  pages = {5888--5891},
  numpages = {0},
  year = {1998},
  month = {Dec},
  publisher = {American Physical Society},
  doi = {10.1103/PhysRevLett.81.5888},
  url = {https://link.aps.org/doi/10.1103/PhysRevLett.81.5888}
}

@article{PhysRevLett.129.253001,
  title = {Topological Flat Bands in Self-Complementary Plasmonic Metasurfaces},
  author = {Xu, Zhixia and Kong, Xianghong and Chang, Jie and Sievenpiper, Daniel F. and Cui, Tie Jun},
  journal = {Phys. Rev. Lett.},
  volume = {129},
  issue = {25},
  pages = {253001},
  numpages = {5},
  year = {2022},
  month = {Dec},
  publisher = {American Physical Society},
  doi = {10.1103/PhysRevLett.129.253001},
  url = {https://link.aps.org/doi/10.1103/PhysRevLett.129.253001}
}

@article{lape2025realization,
  title = {Realization and characterization of an all-bands-flat electrical lattice},
  author = {Lape, Noah and Diubenkov, Simon and English, L. Q. and Kevrekidis, P. G. and Andreanov, Alexei and Kim, Yeongjun and Flach, Sergej},
  journal = {Phys. Rev. B},
  volume = {112},
  issue = {18},
  pages = {184309},
  numpages = {7},
  year = {2025},
  month = {Nov},
  publisher = {American Physical Society},
  doi = {10.1103/1w5c-nsmh},
  url = {https://link.aps.org/doi/10.1103/1w5c-nsmh}
}

@article{taie2015coherent,
  title={Coherent driving and freezing of bosonic matter wave in an optical Lieb lattice},
  author={Taie, Shintaro and Ozawa, Hideki and Ichinose, Tomohiro and Nishio, Takuei and Nakajima, Shuta and Takahashi, Yoshiro},
  journal={Science Advances},
  volume={1},
  number={10},
  pages={e1500854},
  year={2015},
  publisher={American Association for the Advancement of Science}
}

@article{aidelsburger2015measuring,
  title={Measuring the Chern number of Hofstadter bands with ultracold bosonic atoms},
  author={Aidelsburger, Monika and Lohse, Michael and Schweizer, Christian and Atala, Marcos and Barreiro, Julio T and Nascimb{\`e}ne, Sylvain and Cooper, NR and Bloch, Immanuel and Goldman, Nathan},
  journal={Nature Physics},
  volume={11},
  number={2},
  pages={162--166},
  year={2015},
  publisher={Nature Publishing Group UK London}
}

@article{vicencio2015observation,
  title={Observation of localized states in Lieb photonic lattices},
  author={Vicencio, Rodrigo A and Cantillano, Camilo and Morales-Inostroza, Luis and Real, Basti{\'a}n and Mej{\'\i}a-Cort{\'e}s, Cristian and Weimann, Steffen and Szameit, Alexander and Molina, Mario I},
  journal={Physical review letters},
  volume={114},
  number={24},
  pages={245503},
  year={2015},
  publisher={APS}
}

@article{baboux2016bosonic,
  title={Bosonic condensation and disorder-induced localization in a flat band},
  author={Baboux, F and Ge, L and Jacqmin, Thibault and Biondi, M and Galopin, E and Lema{\^\i}tre, A and Le Gratiet, L and Sagnes, Isabelle and Schmidt, S and T{\"u}reci, Hakan E and others},
  journal={Physical review letters},
  volume={116},
  number={6},
  pages={066402},
  year={2016},
  publisher={APS}
}

@article{klembt2017polariton,
  title={Polariton condensation in S-and P-flatbands in a two-dimensional Lieb lattice},
  author={Klembt, Sebastian and Harder, Tristan H and Egorov, Oleg A and Winkler, Karol and Suchomel, Holger and Beierlein, Johannes and Emmerling, Monika and Schneider, Christian and H{\"o}fling, Sven},
  journal={Applied Physics Letters},
  volume={111},
  number={23},
  year={2017},
  publisher={AIP Publishing}
}

@article{alyatkin2021quantum,
  title={Quantum fluids of light in all-optical scatterer lattices},
  author={Alyatkin, Sergey and Sigurdsson, Helgi and Askitopoulos, Alexis and T{\"o}pfer, Julian D and Lagoudakis, Pavlos G},
  journal={Nature Communications},
  volume={12},
  number={1},
  pages={5571},
  year={2021},
  publisher={Nature Publishing Group UK London}
}

@article{berloff2017realizing,
  title={Realizing the classical XY Hamiltonian in polariton simulators},
  author={Berloff, Natalia G and Silva, Matteo and Kalinin, Kirill and Askitopoulos, Alexis and T{\"o}pfer, Julian D and Cilibrizzi, Pasquale and Langbein, Wolfgang and Lagoudakis, Pavlos G},
  journal={Nature materials},
  volume={16},
  number={11},
  pages={1120--1126},
  year={2017},
  publisher={Nature Publishing Group UK London}
}

@article{nixon2013observing,
  title={Observing geometric frustration with thousands of coupled lasers},
  author={Nixon, Micha and Ronen, Eitan and Friesem, Asher A and Davidson, Nir},
  journal={Physical review letters},
  volume={110},
  number={18},
  pages={184102},
  year={2013},
  publisher={APS}
}

@article{cookson2021geometric,
  title={Geometric frustration in polygons of polariton condensates creating vortices of varying topological charge},
  author={Cookson, Tamsin and Kalinin, Kirill and Sigurdsson, Helgi and T{\"o}pfer, Julian D and Alyatkin, Sergey and Silva, Matteo and Langbein, Wolfgang and Berloff, Natalia G and Lagoudakis, Pavlos G},
  journal={Nature communications},
  volume={12},
  number={1},
  pages={2120},
  year={2021},
  publisher={Nature Publishing Group UK London}
}

@article{johnston2021artificial,
  title={Artificial polariton molecules},
  author={Johnston, Alexander and Kalinin, Kirill P and Berloff, Natalia G},
  journal={Physical Review B},
  volume={103},
  number={6},
  pages={L060507},
  year={2021},
  publisher={APS}
}

@article{dolinina2022spontaneous,
  title={Spontaneous symmetry breaking and the dynamics of three interacting nonlinear optical resonators with gain and loss},
  author={Dolinina, D and Yulin, A},
  journal={Physical Review E},
  volume={105},
  number={3},
  pages={034203},
  year={2022},
  publisher={APS}
}

@article{hsu2016bound,
  title={Bound states in the continuum},
  author={Hsu, Chia Wei and Zhen, Bo and Stone, A Douglas and Joannopoulos, John D and Solja{\v{c}}i{\'c}, Marin},
  journal={Nature Reviews Materials},
  volume={1},
  number={9},
  pages={1--13},
  year={2016},
  publisher={Nature Publishing Group}
}

@article{plotnik2011experimental,
  title={Experimental observation of optical bound states in the continuum},
  author={Plotnik, Yonatan and Peleg, Or and Dreisow, Felix and Heinrich, Matthias and Nolte, Stefan and Szameit, Alexander and Segev, Mordechai},
  journal={Physical review letters},
  volume={107},
  number={18},
  pages={183901},
  year={2011},
  publisher={APS}
}

@article{bulgakov2008bound,
  title={Bound states in the continuum in photonic waveguides inspired by defects},
  author={Bulgakov, Evgeny N and Sadreev, Almas F},
  journal={Physical Review B—Condensed Matter and Materials Physics},
  volume={78},
  number={7},
  pages={075105},
  year={2008},
  publisher={APS}
}

@article{topfer2020time,
  title={Time-delay polaritonics},
  author={T{\"o}pfer, Julian D and Sigurdsson, Helgi and Pickup, Lucinda and Lagoudakis, Pavlos G},
  journal={Communications Physics},
  volume={3},
  number={1},
  pages={2},
  year={2020},
  publisher={Nature Publishing Group UK London}
}

@article{PhysRevLett.99.140402,
  title = {Excitations in a Nonequilibrium Bose-Einstein Condensate of Exciton Polaritons},
  author = {Wouters, Michiel and Carusotto, Iacopo},
  journal = {Phys. Rev. Lett.},
  volume = {99},
  issue = {14},
  pages = {140402},
  numpages = {4},
  year = {2007},
  month = {Oct},
  publisher = {American Physical Society},
  doi = {10.1103/PhysRevLett.99.140402},
  url = {https://link.aps.org/doi/10.1103/PhysRevLett.99.140402}
}

@article{PhysRevA.58.4841,
  title = {Generic model of an atom laser},
  author = {Kneer, B. and Wong, T. and Vogel, K. and Schleich, W. P and Walls, D. F.},
  journal = {Phys. Rev. A},
  volume = {58},
  issue = {6},
  pages = {4841--4853},
  numpages = {0},
  year = {1998},
  month = {Dec},
  publisher = {American Physical Society},
  doi = {10.1103/PhysRevA.58.4841},
  url = {https://link.aps.org/doi/10.1103/PhysRevA.58.4841}
}

@book{kuramoto1984chemical,
  title={Chemical turbulence},
  author={Kuramoto, Yoshiki and Kuramoto, Yoshiki},
  year={1984},
  publisher={Springer}
}

@article{PhysRevB.85.085209,
  title = {Quantum anomalous Hall effect in a flat band ferromagnet},
  author = {Zhao, An and Shen, Shun-Qing},
  journal = {Phys. Rev. B},
  volume = {85},
  issue = {8},
  pages = {085209},
  numpages = {6},
  year = {2012},
  month = {Feb},
  publisher = {American Physical Society},
  doi = {10.1103/PhysRevB.85.085209},
  url = {https://link.aps.org/doi/10.1103/PhysRevB.85.085209}
}

@article{PhysRevB.96.205304,
  title = {Disorder-induced topological phase transitions on Lieb lattices},
  author = {Chen, Rui and Xu, Dong-Hui and Zhou, Bin},
  journal = {Phys. Rev. B},
  volume = {96},
  issue = {20},
  pages = {205304},
  numpages = {7},
  year = {2017},
  month = {Nov},
  publisher = {American Physical Society},
  doi = {10.1103/PhysRevB.96.205304},
  url = {https://link.aps.org/doi/10.1103/PhysRevB.96.205304}
}

@article{tsai2015interaction,
  title={Interaction-driven topological and nematic phases on the Lieb lattice},
  author={Tsai, Wei-Feng and Fang, Chen and Yao, Hong and Hu, Jiangping},
  journal={New Journal of Physics},
  volume={17},
  number={5},
  pages={055016},
  year={2015},
  publisher={IOP Publishing}
}

@article{PhysRevA.88.063613,
  title = {Phase diagram and pair Tomonaga-Luttinger liquid in a Bose-Hubbard model with flat bands},
  author = {Takayoshi, Shintaro and Katsura, Hosho and Watanabe, Noriaki and Aoki, Hideo},
  journal = {Phys. Rev. A},
  volume = {88},
  issue = {6},
  pages = {063613},
  numpages = {5},
  year = {2013},
  month = {Dec},
  publisher = {American Physical Society},
  doi = {10.1103/PhysRevA.88.063613},
  url = {https://link.aps.org/doi/10.1103/PhysRevA.88.063613}
}

@article{PhysRevB.88.224203,
  title = {Flat band states: Disorder and nonlinearity},
  author = {Leykam, Daniel and Flach, Sergej and Bahat-Treidel, Omri and Desyatnikov, Anton S.},
  journal = {Phys. Rev. B},
  volume = {88},
  issue = {22},
  pages = {224203},
  numpages = {6},
  year = {2013},
  month = {Dec},
  publisher = {American Physical Society},
  doi = {10.1103/PhysRevB.88.224203},
  url = {https://link.aps.org/doi/10.1103/PhysRevB.88.224203}
}

@article{johnston2024macroscopic,
  title={Macroscopic noise amplification by asymmetric dyads in non-hermitian optical systems for generative diffusion models},
  author={Johnston, Alexander and Berloff, Natalia G},
  journal={Physical Review Letters},
  volume={132},
  number={9},
  pages={096901},
  year={2024},
  publisher={APS}
}

@article{Krauss2007,
  author  = {T. F. Krauss},
  title   = {Slow light in photonic crystal waveguides},
  journal = {Journal of Physics D: Applied Physics},
  volume  = {40},
  number  = {9},
  pages   = {2666},
  year    = {2007},
  DOI     = {10.1088/0022-3727/40/9/S07}
}

@article{Hirose_2014,
  title={Watt-class high-power, high-beam-quality photonic-crystal lasers},
  author={Hirose, Kazuyoshi and Liang, Yong and Kurosaka, Yoshitaka and Watanabe, Akiyoshi and Sugiyama, Takahiro and Noda, Susumu},
  journal={Nature Photonics},
  volume={8},
  number={5},
  pages={406--411},
  year={2014},
  publisher={Springer Science and Business Media LLC},
  doi={10.1038/nphoton.2014.75}
}

@article{Yanik2011,
  author = {Yanik, A. A. and Huang, M. and Artar, A. and Ciraci, T. and Altug, H.},
  title = {Seeing protein monolayers with naked eye through plasmonic {Fano} resonances},
  journal = {{PNAS}},
  volume = {108},
  number = {29},
  pages = {11784--11789},
  year = {2011},
  doi = {10.1073/pnas.1101910108},
  publisher = {National Academy of Sciences}
}

@incollection{Ramachandran2018,
  title={Fano resonances in flat band networks},
  author={Ramachandran, Ajith and Danieli, Carlo and Flach, Sergej},
  booktitle={Fano Resonances in Optics and Microwaves: Physics and Applications},
  pages={311--329},
  year={2018},
  publisher={Springer}
}

@article{ko2020partial,
  title={Partial quantum revivals of localized condensates in distorted lattices},
  author={Ko, Dogyun and Sun, Meng and Andreanov, Alexei and Rubo, YG and Savenko, IG},
  journal={Optics Letters},
  volume={45},
  number={6},
  pages={1571--1574},
  year={2020},
  publisher={Optical Society of America}
}

@article{Sun2018Excitation,
  title = {Excitation of localized condensates in the flat band of the exciton-polariton Lieb lattice},
  author = {Sun, Meng and Savenko, I. G. and Flach, S. and Rubo, Y. G.},
  journal = {Phys. Rev. B},
  volume = {98},
  issue = {16},
  pages = {161204},
  numpages = {5},
  year = {2018},
  month = {Oct},
  publisher = {American Physical Society},
  doi = {10.1103/PhysRevB.98.161204},
  url = {https://link.aps.org/doi/10.1103/PhysRevB.98.161204}
}

@article{whittaker2018exciton,
  title={Exciton polaritons in a two-dimensional Lieb lattice with spin-orbit coupling},
  author={Whittaker, CE and Cancellieri, Emiliano and Walker, PM and Gulevich, DR and Schomerus, H and Vaitiekus, D and Royall, B and Whittaker, DM and Clarke, E and Iorsh, IV and others},
  journal={Physical review letters},
  volume={120},
  number={9},
  pages={097401},
  year={2018},
  publisher={APS}
}

@article{Aleiner2012Radiative,
  title = {Radiative coupling and weak lasing of exciton-polariton condensates},
  author = {Aleiner, I. L. and Altshuler, B. L. and Rubo, Y. G.},
  journal = {Phys. Rev. B},
  volume = {85},
  issue = {12},
  pages = {121301},
  numpages = {4},
  year = {2012},
  month = {Mar},
  publisher = {American Physical Society},
  doi = {10.1103/PhysRevB.85.121301},
  url = {https://link.aps.org/doi/10.1103/PhysRevB.85.121301}
}

@misc{Sim_parameters,
  note = {The parameters of equation (1) used for numericcal simulations. For results shown in Fig 2: $\sigma = 0.4$, $\gamma = 1$, $g=1.5$, $s=1$, $\omega_0 = 3$, $\eta = 0$. For results shown in Fig 4: $\sigma = 0.4$, $\gamma = 1$, $\Gamma =3$, $g=1.5$, $s=1$, $\omega_0 = -1$, $L\approx 2.2$, $\omega_W = 0$, $\eta = 1.5$, $L_W \approx 2.5$.}
}

@article{Leykam:2018aa,
	author = {Leykam, Daniel and Flach, Sergej},
	isbn = {2378-0967},
	journal = {APL Photonics},
	number = {7},
	pages = {070901},
	publisher = {AIP Publishing},
	title = {Perspective: Photonic flatbands},
	ty = {JOUR},
	volume = {3},
	year = {2018}}

@article{danieli2024flat,
	author = {Carlo Danieli and Alexei Andreanov and Daniel Leykam and Sergej Flach},
	doi = {doi:10.1515/nanoph-2024-0135},
	journal = {Nanophotonics},
	number = {21},
	pages = {3925--3944},
	title = {Flat band fine-tuning and its photonic applications},
	url = {https://doi.org/10.1515/nanoph-2024-0135},
	volume = {13},
	year = {2024}}

@misc{danieli2026progressartificialflatband,
	archiveprefix = {arXiv},
	author = {Carlo Danieli and Sergej Flach},
	eprint = {2603.04248},
	primaryclass = {cond-mat.mes-hall},
	title = {Progress on artificial flat band systems: classifying, perturbing, applying},
	url = {https://arxiv.org/abs/2603.04248},
	year = {2026}}

\end{document}


\title{Supplementary materials for \textit{Synchronization-induced flat bands in driven-dissipative dimer-waveguide chains}}
\author{A.~N.~Osipov}
\affiliation{Technion – Israel Institute of Technology, 32000 Haifa, Israel}
\affiliation{Department of Physics, Guangdong Technion–Israel Institute of Technology, 241 Daxue Road, Shantou, Guangdong 515063, China}
\affiliation{Department of Physics, ITMO University, Saint~Petersburg 197101, Russia}

\author{I.~G.~Savenko}
\affiliation{Department of Physics, Guangdong Technion–Israel Institute of Technology, 241 Daxue Road, Shantou, Guangdong 515063, China}
\affiliation{Technion – Israel Institute of Technology, 32000 Haifa, Israel}

\author{Sergej Flach}
\affiliation{Institute for Basic Science, Daejeon 34126, Republic of Korea}
\affiliation{Basic Science Program, Korea University of Science and Technology, Daejeon 34113, Republic of Korea}
\affiliation{Centre for Theoretical Chemistry and Physics, The New Zealand Institute for Advanced
Study (NZIAS), Massey University Albany, Auckland 0745,
New Zealand}

\author{A.~V.~Yulin}
\affiliation{Department of Physics, ITMO University, Saint~Petersburg 197101, Russia}

\maketitle
\section{Derivation of effective equation for interacting lasing resonators}

We start our derivation by considering a single lasing dimer: two resonators coupled to a common waveguide $A$, described by the following system of equations:
\begin{subequations}
\label{One WG dynamic equations}
\begin{align}
    &\partial_t \psi_i(t) = H_0(\psi_i) + \sigma A(t, x_i),\\
    &\partial_t A(t,x) = -\gamma A + \hat{D}A + \sum_i \sigma\delta(x-x_i)\psi_i.
\end{align}
\end{subequations}
The equation for the slowly varying amplitude of the waveguide mode is linear; therefore, the waveguide amplitude $A$ can be expressed in terms of the resonator order parameters $\psi_{B, C}$ using the Green's function for the waveguide mode:
\begin{equation}
    A(t, x) = \sigma\sum_{i}\int_{0}^{\infty} G(x-x_i, \tau) \psi_i(t-\tau) \, d\tau.
    \label{General expression for A}
\end{equation}
The Green's function $G$ is defined by the governing equation:
\begin{equation}
    \left(\partial_t + \gamma - \frac{i}{2}\frac{\partial^2}{\partial x^2} \right)G(x - x_0, t -t_0) = \delta(x-x_0)\delta(t-t_0).
    \label{G definition}
\end{equation}
In reciprocal $(k, \omega)$ space, the equation for the Green's function takes the form:
\begin{equation}
    -i\omega G(k, \omega) - (-\gamma + i\omega_0)G + \frac{i}{2} k^2 G = 1,
\end{equation}
yielding the solution:
\begin{equation}
    G(k, \omega) = \frac{i}{i\gamma + (\omega - (\omega_0 + \frac{1}{2}k^2))}.
\end{equation}
After applying the inverse Fourier transform, we obtain an expression for the Green's function in the coordinate-time domain:
\begin{equation}
     G(\Delta x, \Delta t) = -\frac{e^{-\gamma \Delta t }}{2\pi}\int_{-\infty}^{\infty} dk \, e^{i(k \Delta x - \frac{k^2}{2} \Delta t)}.
     \label{General Green's function}
\end{equation}
Substituting~\eqref{General expression for A} into~\eqref{One WG dynamic equations} yields equations for coupled condensates featuring retarded interactions. Such retarded responses can induce complex phenomena, including spontaneous symmetry breaking, multistability~\cite{dolinina2022spontaneous}, oscillating solutions~\cite{topfer2020time}, or chaotic behavior.
However, if the interaction between the dimers is weak, the resulting dynamics is slow, allowing us to neglect the photon transit time between dimers within the waveguide. Thus, we treat the waveguide-mediated interactions as quasi-instantaneous. Mathematically, this implies that the perturbation to the condensate frequency is small, enabling the ansatz $\psi_i(t) = e^{i\omega t}\xi_i(t)$, where $\xi_i(t)$ varies slowly. Consequently, $\psi_i(t-\tau) = e^{i\omega (t-\tau)} \sum_n (-\tau)^n\frac{\xi_i^{(n)}(t)}{n!}$ represents a series expansion in terms of the delay time $\tau$. 
Retention of only the first term is justified for small enough couplings $\sigma$ providing slow synchronization dynamics that is much slower than the characteristic delay time for small distances between the spots governing fast excitation propagation.

\begin{figure}[!tb]
\centering
\includegraphics[width=0.5\linewidth]{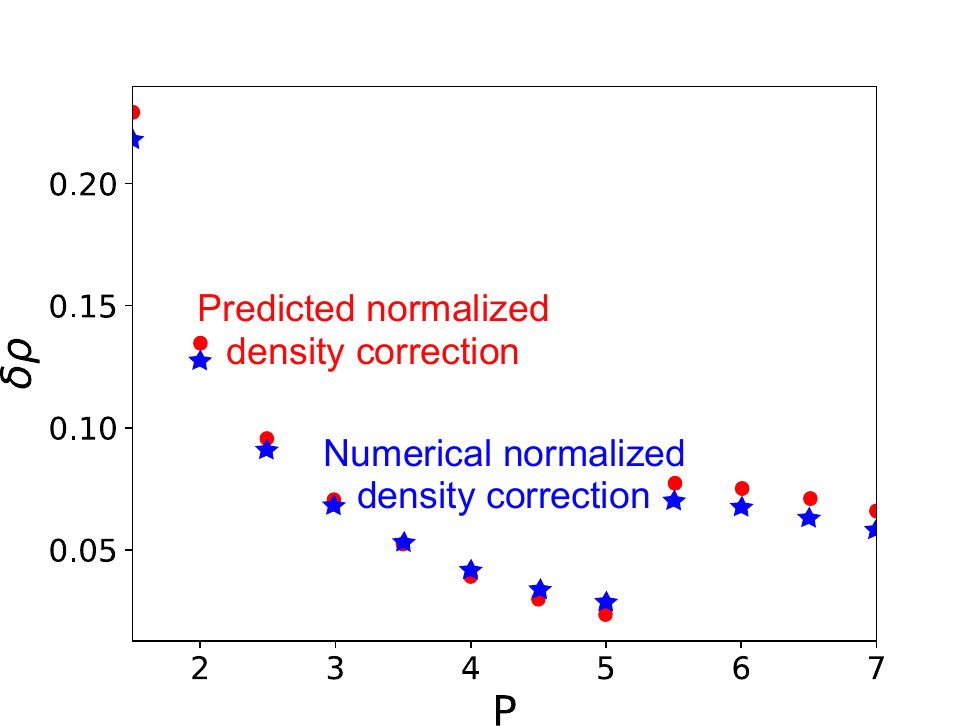}
\caption{Normalized correction of the laser's field density, defined as $\delta \rho = (\rho_{B,C} - (P-1))/(P-1)$, plotted as a function of pump intensity $P$. The analytical prediction (red curve) is given by $\rho_{\text{pr}} = P/(1-\delta \gamma) - 1$, where $\delta\gamma = \sigma^2 \operatorname{Re}[(1+S_1 e^{i\kappa L})/\kappa]$. Blue markers represent $\delta \rho$ obtained from direct numerical simulations of Eq.~\eqref{One WG dynamic equations}. Notably, between $P = 5$ and $P = 5.5$, the resonators undergo a synchronization regime switching from in-phase ($S_1 = 1$) to anti-phase ($S_1 = -1$), which significantly affects the resonator density.}
\label{fig_S1}
\end{figure}

The waveguide mode amplitude at the interaction points of interest reads as:
\begin{equation}
    A(t, x_{B,C}) = \sigma\sum_{i}\int_{0}^{\infty} G(x-x_i, \tau) \psi_i(t-\tau) \, d\tau = \sigma (I_0 + I_L),
    \label{expression for A}
\end{equation}
where the integrals $I_0$ and $I_L$ are defined as follows:
\begin{subequations}
\label{Integrals I0 IL}
\begin{align}
    I_0 &= \frac{1}{2\pi}\int_0^\infty d\tau \, e^{-\gamma \tau - i\omega_0 \tau} \psi_i(t-\tau) \int dk e^{-i \tau k^2/2}, \\
    I_L &= \frac{1}{2\pi}\int_0^\infty d\tau \, e^{-\gamma \tau - i\omega_0 \tau} \psi_j(t-\tau) \int dk e^{ikL - i \tau k^2/2}.
\end{align}
\end{subequations}
As previously noted, the integration can be performed by substituting the ansatz $\psi_i(t) = e^{i\omega t}\xi_i(t)$, where $\omega$ is the unperturbed condensate frequency satisfying $-i\omega \psi = H_0(\psi)$. Integrating over the Green's function poles under the condition $\operatorname{Im}[\sqrt{2(\omega-\omega_0) + 2i\gamma}]>0$ we obtain:
\begin{subequations}
\label{AnalyticalIntegrals}
\begin{align}
    I_0 &= \frac{1}{2\pi}\psi_i(t)\int_{-\infty}^{+\infty} \frac{dk}{i(\omega_0 - \omega)+ \gamma + ik^2/2 } = \frac{\psi_i(t)}{\sqrt{2(\omega-\omega_0) + 2i\gamma}}, \\
    I_L &= \frac{1}{2\pi}\psi_j(t)\int_{-\infty}^{+\infty} \frac{e^{ik L} \, dk}{i(\omega_0 - \omega)+ \gamma + ik^2/2 } = \psi_j(t)\frac{e^{i L \sqrt{2(\omega-\omega_0) + 2i\gamma}}}{\sqrt{2(\omega-\omega_0) + 2i\gamma}}.
\end{align}
\end{subequations}
By introducing the effective wave vector of the waveguide mode, $\kappa = \sqrt{2(\omega-\omega_0) + 2i\gamma}$, we arrive at:
\begin{eqnarray}
    I_0 = \frac{\psi_{B,C}(t)}{\kappa}, \quad I_L = \psi_{C,B}(t)\frac{e^{i \kappa L }}{\kappa},
    \label{coupling integral}
\end{eqnarray}
where $i = B,C$ corresponds to the self-energy shift and $j = C, B$ to the coupling between resonators. Substituting these expressions into Eq.~\eqref{expression for A} and subsequently into Eq.~\eqref{One WG dynamic equations}, we obtain the effective coupled-mode equations:
\begin{equation}
    \partial_t \psi_{B, C}(t) = H_0(\psi_{B, C}) + \frac{\sigma^2}{\kappa}\psi_{B, C} + \frac{\sigma^2 e^{i\kappa L}}{\kappa}\psi_{C, B}.
    \label{Coupled modes eq 0}
\end{equation}

In addition to the synchronization regimes discussed in the main text, we compare here the density modification induced by coupling to waveguide $A$, obtained from direct numerical simulations, with our analytical predictions based on Eq.~\eqref{Coupled modes eq 0}. Figure~\ref{fig_S1} demonstrates excellent agreement between the numerically simulated and analytically calculated corrections to the resonator field density $\rho$, confirming that our model accurately predicts not only the sign but also the magnitude of the coupling.
%
%
%
%
\section{Derivation of effective equations for the chain of coupled laser dimers}

Following the approach used for a single dimer, we begin the derivation for the dimer chain by expressing the waveguide amplitudes in Eq.~(1) of the main text using the respective Green's functions:
\begin{equation}
    A(t, x) = \sigma\sum_{a}\int G_A(x-x_a, \tau) \delta(x_a)\psi_a(t-\tau) + \eta \int G_A(x-x_0, \tau) \delta(x_0)W(t-\tau, y_i),
    \label{A expression}
\end{equation}
\begin{equation}
    W(t, y) = \eta\sum_{i}\int G_W(y-y_i, \tau) \delta(y_i)A_i(t-\tau, x_0).
    \label{W expression}
\end{equation}
Substituting Eq.~\eqref{W expression} into Eq.~\eqref{A expression}, we obtain:
\begin{eqnarray}
    A_i(t, x) &=& \sigma\sum_{a}\int G_A(x-x_a, \tau) \delta(x_a)\psi_a(t-\tau)\\
    \nonumber
    &&+ \eta^2 \sum_j \int dx d\tau_1 G_A(x-x_0, \tau_1) \delta(x_0)\int dy d\tau_2G_W(y-y_j, \tau_2) \delta(y_j)A_j(t-\tau_1-\tau_2, x_0).
\end{eqnarray}
In the limit of weak coupling between dimers, we treat the interaction with waveguide $W$ as a perturbation. This allows us to replace the amplitude $A$ in the double integral with the unperturbed amplitude:
\begin{equation}
    A_i^{(0)}(t, x_0) = \sigma(\psi_{B_i}(t) + \psi_{C_i}(t))\frac{e^{i\kappa L/2}}{\kappa}.
\end{equation}
Using the same integration procedure as in the previous section and neglecting time-delay effects we arrive at the final expression for the waveguide amplitude:
\begin{equation}
    A_i(t, x) = \sigma\left(\psi_{B_i }\frac{e^{i\kappa|x - x_{B_i}|} }{\kappa} + \psi_{C_i }\frac{e^{i\kappa|x - x_{C_i}|} }{\kappa}\right) + \eta^2 \sigma \sum_j \frac{e^{i\kappa_W |y_i-y_j|}}{\kappa_W}\frac{e^{i\kappa|x - x_0|} }{\kappa}\frac{e^{i\kappa L/2}}{\kappa} (\psi_{B_i}(t) + \psi_{C_i}(t)),
    \label{expression for A final}
\end{equation}
where $\kappa_W = \sqrt{2(\omega-\omega_W) + 2i\Gamma}$ and $\kappa = \sqrt{2(\omega-\omega_0) + 2i\gamma}$ represent the effective wave vectors in the horizontal and vertical waveguides on the frequency of uncoupled resonator's $\omega$, respectively. 
Substituting Eq.~\eqref{expression for A final} back into the dynamical equations for the condensates [Eq.~(1a) from the main text], we obtain the coupled-mode equations for the dimer chain:
\begin{subequations}
\label{Copuling modes equations lattice}
\begin{align}
    \partial_t \psi_{B_i} = H_0(\psi_{B_i}) + \sigma^2\psi_{B_i}/\kappa + \sigma^2\psi_{C_i}e^{i\kappa L}/\kappa + \eta^2\sigma^2\frac{e^{i\kappa L}}{\kappa^2}\sum_{j=0}^N\frac{(e^{i\kappa_W L_W})^{|i-j|}}{\kappa_W}(\psi_{B_j} +\psi_{C_j}),  \\
    \partial_t \psi_{C_i} = H_0(\psi_{C_i}) + \sigma^2\psi_{C_i}/\kappa + \sigma^2\psi_{B_i}e^{i\kappa L}/\kappa + \eta^2\sigma^2\frac{e^{i\kappa L}}{\kappa^2}\sum_{j=0}^N\frac{(e^{i\kappa_W L_W})^{|i-j|}}{\kappa_W}(\psi_{B_j} +\psi_{C_j}), 
\end{align}
\end{subequations}
where $L_W$ is the inter-dimer distance, and $N$ is the total number of dimers in the chain. By assuming that only nearest-neighbor interactions are significant which is valid for sufficiently large inter-dimers distances, we obtain Eq.~(2) with the effective inter-dimer coupling parameters $\zeta = \operatorname{Re}\left[\eta^2\sigma^2\frac{e^{i\kappa L}}{\kappa^2}\frac{e^{i\kappa_W L_W}}{\kappa_W}\right]$ and $\chi = -\operatorname{Im}\left[\eta^2\sigma^2\frac{e^{i\kappa L}}{\kappa^2}\frac{e^{i\kappa_W L_W}}{\kappa_W}\right]/\zeta$.

\section{Derivation of Kuramoto-like approximation}

To provide a detailed description of the synchronization of two coupled resonators, we derive an equation based on the Kuramoto-like approximation~\cite{kuramoto1984chemical}. 
We begin by substituting the density-phase ansatz, $\psi_{B, C} = \sqrt{\rho_{B,C}}e^{i\phi_{B,C}}$, into Eq.~(3) from the main text. 
This yields the dynamical equations for the densities and phases of the resonator order parameters:
\begin{subequations}
\label{eq_density_phase}
\begin{align}
\label{eq density}
&\partial_t \sqrt{\rho_{B,C}} = \left(\frac{P}{1+\rho_{B,C}}-1\right)\sqrt{\rho_{B,C}} + J\left[g\sin(\phi_{C,B}-\phi_{B,C}) +\cos(\phi_{C,B} - \phi_{B,C})\right]\sqrt{\rho_{C,B}},\\
& \sqrt{\rho_{B,C}}\partial_t \phi_{B, C} = -(\kappa \frac{P}{1+\rho_{B, C}} + \alpha\rho_{B,C})\sqrt{\rho_{B,C}} + J\left[-g\cos(\phi_{B, C}-\phi_{C, B}) +\sin(\phi_{C, B}-\phi_{B, C})\right]\sqrt{\rho_{C,B}}.
\label{eq phase}
\end{align}
\end{subequations}
Assuming fast density relaxation to equilibrium, we set $\partial_t\sqrt{\rho_{B,C}} = 0$ and expand the density in a perturbation series: $\sqrt{\rho_{B,C}} = \sqrt{P-1} + \varepsilon^{(1)}_{B,C} + \varepsilon^{(2)}_{B,C}$, where $\varepsilon^{(1)}$ and $\varepsilon^{(2)}$ are the first- and second-order corrections with respect to the coupling strength $J$. Substituting this expansion into Eq. \eqref{eq density}, we find:
\begin{subequations}
\label{epsilons}
\begin{align}
    \varepsilon^{(1)}_{B, C} &= J\frac{P}{2\sqrt{P-1}}\left[\cos(\phi_{B,C}-\phi_{C,B}) - g \sin(\phi_{B,C}-\phi_{C,B})\right], \\
    \varepsilon^{(2)}_{B, C} &= \frac{P}{2(P-1)}\left[J(\cos(\phi_{B,C}-\phi_{C,B}) - g \sin(\phi_{B,C}-\phi_{C,B}))\varepsilon^{(1)}_{B, C} - (\varepsilon^{(1)}_{B, C})^2\frac{\sqrt{P-1}}{P}\left(3 - \frac{4(P-1)}{P}\right) \right].
\end{align}
\end{subequations}

Similarly, expanding the density in the phase equation~\eqref{eq phase} yields:
\begin{align}
    \partial_t \phi_{B, C} = &-s(P-1) -\kappa -\kappa\left(-2\frac{\sqrt{P-1}}{P}\varepsilon^{(2)}_{B, C} - 2\frac{\sqrt{P-1}}{P}\varepsilon^{(1)}_{B, C} - \frac{1}{P}(\varepsilon^{(1)}_{B, C})^2 + \frac{4 (P-1)}{P^2}(\varepsilon^{(1)}_{B, C})^2\right) \nonumber \\
    &-\alpha\left(2\sqrt{P-1}(\varepsilon^{(1)}_{B, C} + \varepsilon^{(2)}_{B, C}) + (\varepsilon^{(1)}_{B, C})^2\right) + J\left(-g\cos(\phi_{B, C}-\phi_{C, B}) +\sin(\phi_{C, B} - \phi_{B, C})\right) \nonumber \\
    &+ J\left(-g\cos(\phi_{B, C}-\phi_{C, B}) +\sin(\phi_{C, B} - \phi_{B, C})\right)\frac{(\varepsilon^{(1)}_{C, B} - \varepsilon^{(1)}_{B, C})}{\sqrt{P-1}}.
    \label{phase expansion}
\end{align}

Finally, substituting Eq. \eqref{epsilons} into Eq. \eqref{phase expansion} and defining the phase difference $\Delta \phi = \phi_B - \phi_C$, we obtain:
\begin{equation}
\label{perturbative expansion2}
    \partial_t (\Delta \phi) = -J(1-s \varkappa P + g \varkappa) \sin(\Delta \phi) - J^2 \frac{\varkappa P\left( \varkappa - 2(P-1)s - g\frac{3P-4}{2P}\right)}{(P-1)}\sin(2\Delta \phi).
\end{equation}
%
%
The term proportional to $J$ here corresponds to the first-order approximation. 
In the regime, when we neglect the terms proportional to $J^2$, the system possesses two equilibrium points: $\Delta \phi = 0$ and $\Delta \phi = \pi$. 
The stability of these solutions is determined by the sign of the effective coupling; specifically, a solution $\Delta \phi_s$ is stable if $J(1-\alpha \varkappa P + g \varkappa) \cos(\Delta \phi_s) > 0$. 

The second-order term proportional to $J^2$ becomes significant as the coupling strength increases, or when the first-order coefficient $1-\alpha \varkappa P + g \varkappa$ vanishes. 
In such cases, the system can support four equilibrium points, with a pair of asymmetric solutions bifurcating from the symmetric states via a pitchfork bifurcation. 
Such spontaneous symmetry breaking has been previously predicted in polaritonic molecules~\cite{johnston2021artificial}.

\begin{figure}[!tb]
\includegraphics[width=0.8\linewidth]{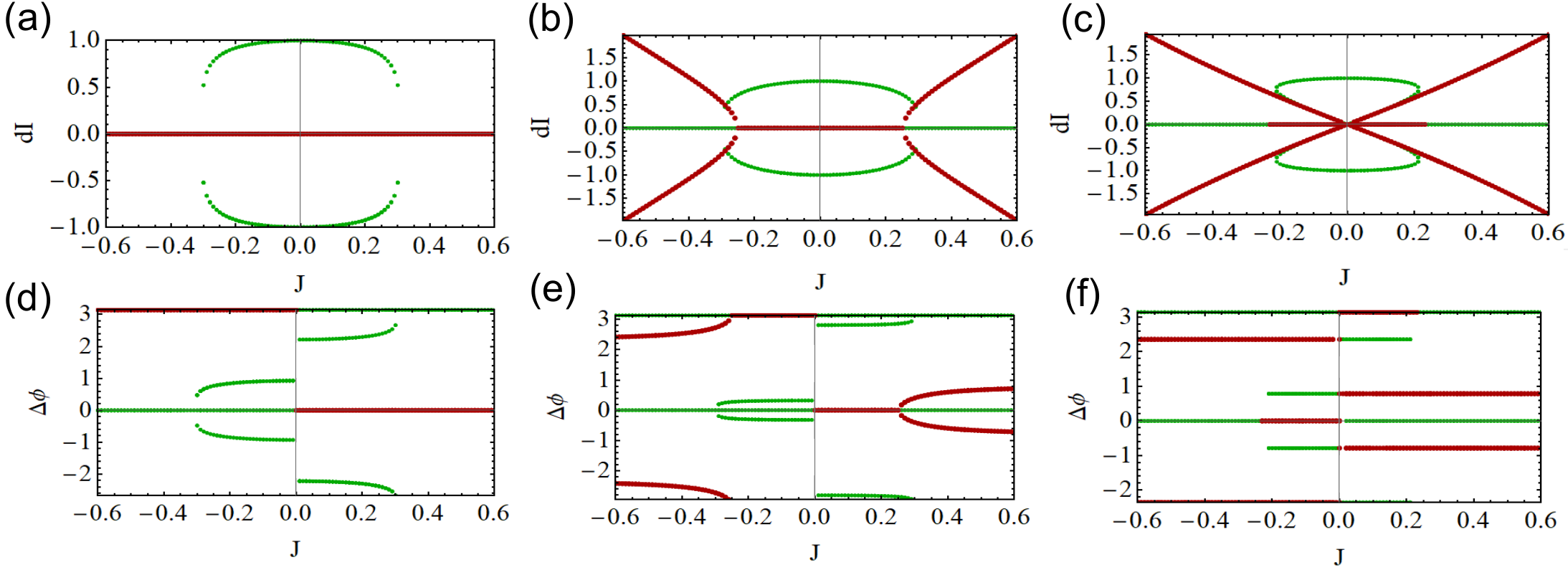}
\caption{(a)-(c) The population imbalance $dI = \rho_B -\rho_C$ calculated from Eq.(3) from the main text as a function of the coupling strength $J$. (d)-(f) The corresponding phase difference $\Delta \phi = \operatorname{Arg}(\psi_B\psi_C^*)$ as a function of $J$. Red curves denote stable solutions; green curves denote unstable ones. Parameters: $P=2$, $s=1$; $g = 1.5$, $\varkappa = -0.5$ for (a,d); $g = 1.5$, $\varkappa = 1$ for (b,e); $g = 1$, $\varkappa = 1$ for (c,f).}
\label{fig_S2}
\end{figure}

We find the equilibrium points of Eq.~(3) from the main text numerically and study their stability. 
Figures~\ref{fig_S2}(a) and (d) illustrate the standard synchronization regime where phase symmetry is determined by the sign of $J_{\text{eff}}$. Panels (b) and (e) show the regime where $J_{\text{eff}}$ is small. 
This leads to the emergence of broken-symmetry states at higher coupling strengths. 
Panels (c) and (f) depict the degenerate case where $1-\alpha \varkappa P + g \varkappa = 0$, causing a symmetry-breaking bifurcation to emerge from the origin point, resulting in bistability between symmetric and asymmetric states.

It is important to note that the sign of the second-order correction in Eq. \eqref{perturbative expansion2} is crucial for the stability of these asymmetric solutions. In our model, this coefficient can be positive or negative. Stable symmetry-breaking states appear only when the second-order coefficient is positive. Conversely, if it is negative, only symmetric states remain stable, as seen in panels (a) and (d). For systems with pure Kerr (cubic) nonlinearity, this term is consistently negative. This can be recovered from our theory by neglecting terms with $P^2$ and $P(P-1)$ in the numerator, yielding $\Delta \phi^{(2)} = -J^2 \frac{P\varkappa^2}{P-1}$, which is consistent with the results typical for low-pumping expansions of Kerr-like media.


\section{Spectra of linear excitations}
To study the spectra of linear excitations for a nonlinear dynamical system of the form
\begin{eqnarray}
    \partial_t\Psi = H(\Psi),
    \label{General eq}
\end{eqnarray}
we substitute the ansatz $\Psi = (\Psi^{(0)} + \xi(t))e^{-i\omega_0 t}$, where $\Psi^{(0)}$ is a known nonlinear solution of Eq.~\eqref{General eq} with frequency $\omega_0$ satisfying $-i\omega_0 \Psi^{(0)} = H(\Psi^{(0)})$, and $\xi$ being a small correction to this stationary solution. Linearizing Eq.~\eqref{General eq} around $\Psi^{(0)}$ yields a system of equations for the correction $\xi$: 
\begin{subequations}
\begin{equation}
    \partial_t \xi = i\omega_0 \xi + \left(\frac{\partial H}{\partial \Psi}\Bigg |_{\Psi^{(0))}}\right)_{(1)} \xi +\left(\frac{\partial H}{\partial \Psi}\Bigg |_{\Psi^{(0))}}\right)_{(2)}\xi^*, 
\end{equation}
\begin{equation}
    \partial_t \xi^* = -i\omega_0 \xi^* + \left(\frac{\partial H}{\partial \Psi}\Bigg |_{\Psi^{(0))}}\right)_{(1)}^{*} \xi^* +\left(\frac{\partial H}{\partial \Psi}\Bigg |_{\Psi^{(0))}}\right)_{(2)}^*\xi, 
\end{equation}
\end{subequations}
where $\left(\frac{\partial H}{\partial \Psi}\Big |_{\Psi^{(0))}}\right)_{(1)}$ and $\left(\frac{\partial H}{\partial \Psi}\Big |_{\Psi^{(0))}}\right)_{(2)}$ denote the components of the linear expansion proportional to $\xi$ and $\xi^*$, respectively. This system can be rewritten in matrix form as
\begin{equation}
    \partial_t \hat{\xi} = \mathcal{L}\hat{\xi},
\end{equation}
with
\begin{equation}
    \mathcal{L} = \begin{pmatrix}i\omega_0 + \left(\frac{\partial H}{\partial \Psi}\Big |_{\Psi^{(0))}}\right)_{(1)} & \left(\frac{\partial H}{\partial \Psi}\Big |_{\Psi^{(0))}}\right)_{(2)} \\
    \left(\frac{\partial H}{\partial \Psi}\Big |_{\Psi^{(0))}}\right)_{(2)}^* & -i\omega_0 + \left(\frac{\partial H}{\partial \Psi}\Big |_{\Psi^{(0))}}\right)_{(1)}^* 
    \end{pmatrix}.
\end{equation}

For the chain of laser dimers with periodic boundary conditions, we can construct explicit plane-wave solutions. For the chain of in-phase phase-locked dimers ($\Delta \phi_i = \phi_{B_i}-\phi_{C_i} = 0$), the condensate intensities and frequencies are given by
\begin{subequations}
\begin{equation}
    \forall \,ijm:\: B_i = B_j = C_m  \rightarrow |B|^2 = |C|^2 = I_0 =\frac{P}{1 - J - 4\zeta} - 1,\quad w = \alpha I_0 + \frac{gP}{1+I_0} +J\varkappa + 4\zeta \chi,
\end{equation}
\begin{equation}
    \forall \,i:\: B_i = C_i = (-1)^iB_0  \rightarrow |B|^2 = |C|^2 = I_0 =\frac{P}{1 - J + 4\zeta} - 1,\quad w = \alpha I_0 + \frac{gP}{1+I_0} +J\varkappa- 4\zeta \chi ,
\end{equation} 
\end{subequations}
for ordered chain configurations with inter-dimer phase differences $0$ and $\pi$ between neighboring dimers, respectively. Note that the antiferromagnetic phase ordering of dimers (with $\pi$ phase difference between neighboring dimers) along the chain requires an even number of dimers; for an odd number of dimers, geometric frustration may arise~\cite{cookson2021geometric}.
\begin{figure}[!tb]
\centering
\includegraphics[width=0.6\linewidth]{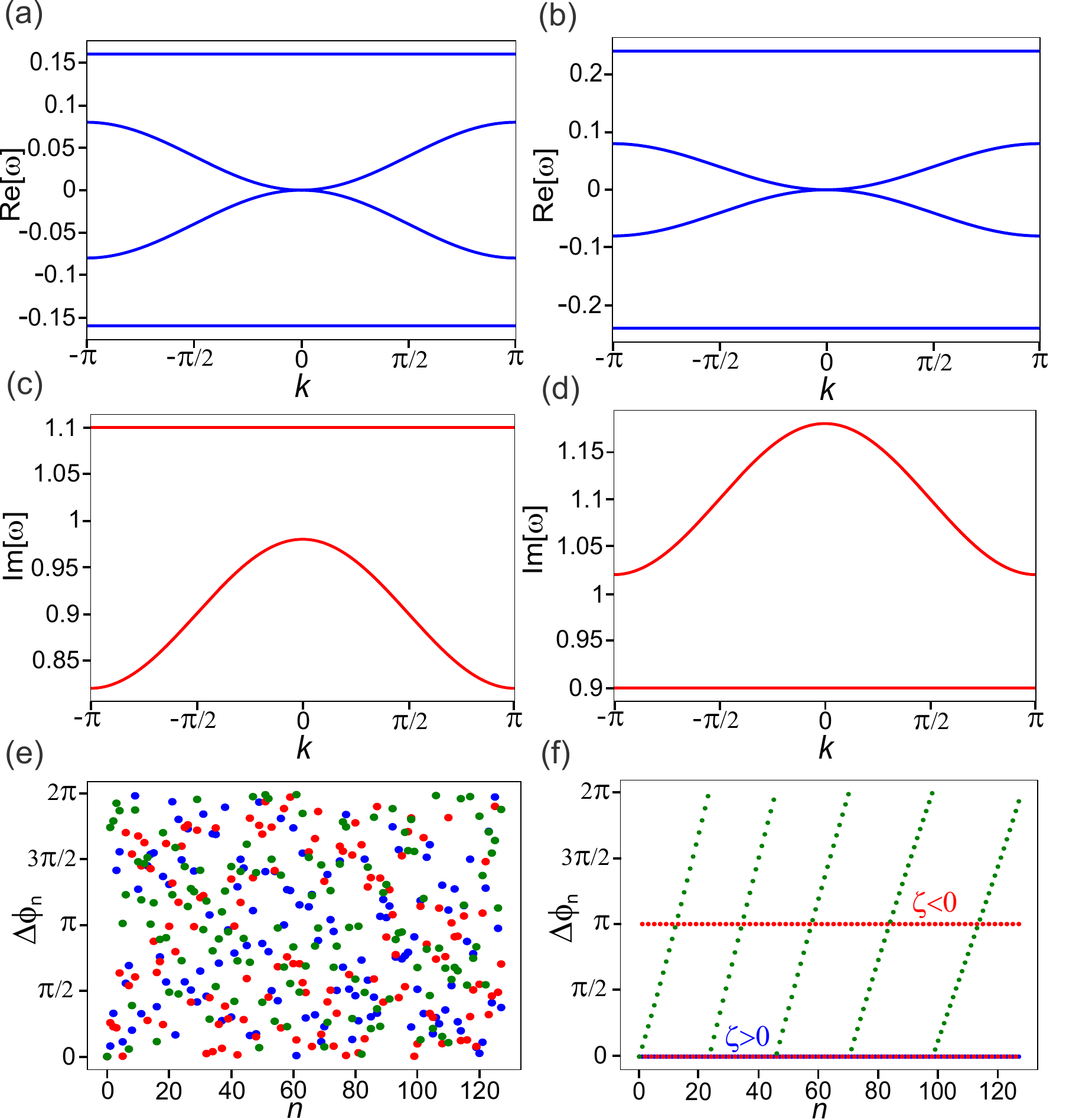}
\caption{Linear excitation spectra of the chain described by Eq.~(2) from the main text. (a,b) Real part of the eigenfrequency as a function of the wave number $k$. (c,d) Corresponding imaginary part (effective gain/damping) as a function of $k$. Panels (a) and (c) show spectra for $J = -0.1$, corresponding to the trivial solution $\psi_{B,C}^{(0)} = 0$; panels (b) and (d) show spectra for $J = 0.1$. (e) Phase distribution of dimers along the chain for noninteracting antisymmetric dimers obtained from simulations starting from random noise. Different colors correspond to different noise realizations. (f) Phase distributions for phase-locked symmetric dimers. Blue and red dots show stable configurations for $\zeta > 0$ (phase difference $0$ between neighboring dimers in chain) and $\zeta < 0$ (phase difference $\pi$ between neighboring dimers in chain), respectively. Blue color corresponds to a stable state with a linear phase gradient along the chain ($\zeta > 0$). Parameters used throughout: $g=1$, $\alpha = 1$, $\kappa = -1$, $|\zeta| = 0.02$, $\chi = -0.5$. For all panels except (b) and (d), $P = 2$.}
\label{fig_S3}
\end{figure}
For anti-phase phase-locked dimers (antisymmetric within each dimer), the solution density and frequency are
\begin{equation}
    \forall \,i:\: B_i = -C_i  \rightarrow |B|^2 = |C|^2 = I_0 =\frac{P}{1 + J} - 1,\quad w = \alpha I_0 + \frac{gP}{1+I_0} -J\varkappa .
\end{equation}
To proceed further, we require translational symmetry of the background solutions. For a ordered phase-locked chain of ferromagnetic dimers, this symmetry is naturally provided by the synchronization between dimers. For antiferromagnetic uncoupled dimers, translational symmetry is imposed manually (see the following section). The perturbation ansatz can then be written as
\begin{equation}
    B_j,C_j = (B_0 + b^r_j + i b^i_j, C_0 + c^r_j + ic^i_j)e^{-iw_0t}.
\end{equation}
Applying the Bloch theorem, we introduce the Fourier transform for the linear corrections: 
\begin{equation}
    [b_n, c_n] = \sum_k  [b_i, c_i] e^{ik n L_{W}}.
\end{equation}
In reciprocal space, the matrix governing the linear dynamics of the corrections takes the form

\begin{align}
    &\mathcal{L}_k =\begin{pmatrix} l_1(B_0) -1  & l_2(B_0) & l_3 & 0 \\
   l_2^*(B_0)& l_1^*(B_0)-1 &0 &l_3^*  \\
    l_3 & 0 & l_1(C_0) &  l_2(C_0) \\
    0 & l_3^*  &  l_2^*(C_0)& l_1^*(C_0)
    \end{pmatrix},
    \label{LE dispersion matrix}
\end{align}

where
\begin{subequations}    
\begin{align*}
&l_1(X) = i\omega + (1-ig)c_1 - 2i\alpha |X|^2 + 2\cos(k)\zeta(1-i\chi), \\
&l_2(X) = (1-ig)c_2(X) - i\alpha X^2, \\
&l_3 = J(1-i\varkappa) + 2\cos(k)\zeta(1-i\chi),
\end{align*}
\end{subequations}
with $c_1 = P/(1+I_0)^2$ and $c_2(X) = P X^2/(1+I_0)^2$. The eigenvalues of $\mathcal{L}_k$ yielding the linear excitation spectra are discussed in the main text.

In addition to spectra of excitations shown in the main text, here we study in more details linear excitations on top of the trivial solution $\psi_{B,C}^{(0)} = 0$ with frequency $\omega_0 = gP + |J|\varkappa + 2(1+\operatorname{sign}(J))|\zeta|\chi$ under pump intensity $P=2$. The resulting spectra are presented in Fig.~\ref{fig_S3}: panels (a) and (c) correspond to $J = -0.1$, while (b) and (d) correspond to $J = 0.1$. For $J = -0.1$, the flat band exhibits the highest linear gain; conversely, for $J = 0.1$, the flat band has the lowest gain. Consequently, when the system evolves from random initial noise to a stable state, $J = -0.1$ yields a random phase distribution characteristic to noninteracting dimers (Fig.~\ref{fig_S3}(e)), whereas $J = 0.1$ leads to various phase-locked ordered configurations (Fig.~\ref{fig_S3}(f)).

\section{Dispersion of linear excitations for the chain of non-interacting dimers for arbitrary phases of dimers}

Consider a nonlinear solution formed at some time $t = t_0$ corresponding to a chain of antiferromagnetic dimers with random phases: $\psi_{B_j,C_j}(t_0) = \pm |\psi_{B_j,C_j}| e^{i\phi_j}$. To study the linear excitations on top of this background, we write the perturbed field as $\psi_{B_j,C_j} = (\pm|\psi_{B_j,C_j}| + \xi_{B_j,C_j}) e^{i\phi_j}$, where $\xi_{B_j,C_j}$ are small corrections. 
Linearizing the equations of motion yields
\begin{eqnarray}
    \partial_t \xi_j^{B,C} e^{i\phi_j} &=& i\omega_0 \xi_j e^{i\phi_j} + \left(\frac{\partial H_0}{\partial \Psi}\Bigg |_{\Psi^{(0))}}\right)_{(1)} \xi_j^{B,C}  e^{i\phi_j} +\left(\frac{\partial H_0}{\partial \Psi}\Bigg |_{\Psi^{(0))}}\right)_{(2)}{\xi_j^{B,C}}^* e^{i\phi_j} \\ \nonumber
    &&+ \zeta(1-i\chi)((\xi_{j-1}^{B} + \xi_{j-1}^{C}) e^{i\phi_{j-1}}  + (\xi_{j+1}^{B} + \xi_{j+1}^{C}) e^{i\phi_{j+1}}).
\end{eqnarray}

For perturbations corresponding to the flat band of Goldstone modes, the analysis simplifies considerably. 
In this case $\xi_i^B = -\xi_i^C$, and we can apply a phase rotation by multiplying the equation for $\xi_i$ by $e^{-i\phi_i}$, recovering the same result as for the homogeneous background. For more general perturbations, we find
\begin{equation}
    \partial_t \xi_j^{B,C} = i\omega_0 \xi_j + \left(\frac{\partial H_0}{\partial \Psi}\Bigg |_{\Psi^{(0))}}\right)_{(1)} \xi_j^{B,C}   +\left(\frac{\partial H_0}{\partial \Psi}\Bigg |_{\Psi^{(0))}}\right)_{(2)}{\xi_j^{B,C}}^* + \zeta_{j-}(\xi_{j-1}^{B} + \xi_{j-1}^{C})  + \zeta_{j+}(\xi_{j+1}^{B} + \xi_{j+1}^{C})  
\end{equation}
where $\zeta_{j\pm} = \zeta(1-i\chi)e^{i(\phi_{j\pm1} - \phi_j)}$. 

Thus, in the case of random phases acquired by individual dimers, the flat band observed in the linear excitation spectra remains a meaningful solution, and it can still be detected in the noise response spectra. 
The dispersive bands, however, are not directly observable in systems lacking translational invariance. Instead, we anticipate the formation of a band of randomly distributed eigenlevels, with energies clustered around the dispersive bands discussed in the main text. 
This makes the question of the system's response to external propagating signals tricky and intriguing.  

In addition, we consider a specific class of phase distributions that preserve translational structure: a linear phase gradient along the chain of the form $\psi_j(k_0) = |\psi| e^{i j k_0}$, with $k_0 = \pi m / N$ ($N$ being the number of dimers). In this case, $\phi_j = \phi_{j-1} + \pi k_0$, and consequently $\zeta_{j\pm} = 2\zeta(1-i\chi) e^{\pm i k_0}$.

Following the same linearization procedure described in the previous section, such a periodic phase modulation modifies the dispersion relation by shifting the wave vector. Specifically, the terms containing $\cos(k)$ in the matrix $\mathcal{L}_k$ of Eq.~\eqref{LE dispersion matrix} are replaced by $\cos(k + k_0)$.
Therefore, a periodic modulation of the phase along the chain of noninteracting dimers does not affect the flat bands but simply shifts the dispersive branches shown in Figs.~3(a) and (c) from the main text by the modulation wave vector $k_0$, without altering their shape.


\bibliographystyle{apsrev4-2}
\bibliography{references}